\documentclass[10pt,journal,final]{IEEEtran}
\usepackage{mathrsfs}
\usepackage{graphicx}
\graphicspath{{./figure_box/}}
\usepackage{float}
\usepackage{booktabs}
\usepackage{threeparttable}
\usepackage{hyperref}
\usepackage{cite}

\usepackage{enumitem}
\usepackage{color}
\usepackage{caption}
\usepackage{psfrag}
\usepackage{subfigure}
\usepackage{amssymb}
\usepackage{amsthm}
\usepackage{setspace}
\usepackage{epsfig}
\usepackage{pifont}
\usepackage{epsfig}
\usepackage{pifont}
\usepackage{amsmath}
\usepackage{array}
\usepackage{multicol}
\usepackage{multirow}
\usepackage{diagbox}
\usepackage{pifont}
\usepackage{indentfirst}
\usepackage{amsfonts}
\usepackage{algorithm}
\usepackage{algorithmicx}
\usepackage{algpseudocode}
\usepackage{fancyhdr}
\usepackage{amscd}
\usepackage{bm}
\usepackage{fancyhdr}
\usepackage{enumerate}
\usepackage{color}
\usepackage{threeparttable}

\newcommand{\T}{\mathsf{T}}
\renewcommand{\H}{\mathsf{H}}

\IEEEoverridecommandlockouts
\makeatletter
\renewcommand{\citepunct}{,\penalty\@m\hskip.13emplus.1emminus.1em}
\renewcommand{\citedash}{\hbox{--}\penalty\@m}

\begin{document}
	\title{Association-Aware GNN for Precoder Learning in Cell-Free Systems}
	
\author{\IEEEauthorblockN{Mingyu Deng and Shengqian Han} \\ 
	\IEEEauthorblockA{School of Electronics and Information Engineering, Beihang University, Beijing 100191, China\\
	Email: \{mingyudeng, sqhan\}@buaa.edu.cn}
    \vspace{-10mm}
}

\maketitle

\begin{abstract}
Deep learning has been widely recognized as a promising approach for optimizing multi-user multi-antenna precoders in traditional cellular systems. However, a critical distinction between cell-free and cellular systems lies in the flexibility of user equipment (UE)-access point (AP) associations. Consequently, the optimal precoder depends not only on channel state information but also on the dynamic UE–AP association status. In this paper, we propose an association-aware graph neural network (AAGNN) that explicitly incorporates association status into the precoding design. We leverage the permutation equivariance properties of the cell-free precoding policy to reduce the training complexity of AAGNN and employ an attention mechanism to enhance its generalization performance. Simulation results demonstrate that the proposed AAGNN outperforms baseline learning methods in both learning performance and generalization capabilities while maintaining low training and inference complexity.
\end{abstract}

\begin{IEEEkeywords}
Cell-free, precoding, graph neural network, permutation equivariance
\vspace{-3mm}
\end{IEEEkeywords}

\section{Introduction}
Cell-free multi-antenna systems effectively mitigate inter-cell interference by jointly serving user equipments (UEs) via coordinated precoding across adjacent access points (APs)~\cite{Demir2021Cellfree}. However, the joint optimization required for this coordination introduces significant computational overhead. Although various numerical algorithms have been developed for precoder optimization~\cite{Demir2021Cellfree}, their high complexity makes it challenging to meet the real-time requirements of practical systems.

In recent years, deep learning methods have been widely used to learn precoding policies in cellular networks to reduce online inference latency \cite{Shi2018DNN, Kim2020LearningMethod}. For example, fully-connected neural networks (FNNs) can achieve near-optimal performance with low inference complexity \cite{Shi2018DNN}. Convolutional neural networks (CNNs), originally designed to exploit the shift-invariant property of  image processing tasks \cite{Bronstein2021InductiveBiase}, also perform well in precoder learning \cite{Kim2020LearningMethod}. Compared to these conventional deep neural networks (DNNs), graph neural networks (GNNs) typically offer superior learning performance, lower training complexity, and better generalizability across varying system sizes \cite{Zhao2022EGNN}. These advantages stem from the permutation equivariance (PE) properties of GNNs, which inherently align with the PE properties of precoding policies.

Precoding policies in cell-free systems exhibit a 3D-PE property with respect to UEs, APs, and AP antennas. Specifically, if the indices of UEs, APs, or antennas in the channel matrix are permuted, the corresponding indices in the output precoder matrix must be permuted accordingly. However, existing DNNs for cell-free precoding have not fully exploited this property. For instance, the FNN and CNN architectures in \cite{Aggarwal2023CellfreeDNN} and \cite{Chen2024SUNet} do not satisfy any PE properties, while the heterogeneous GNN in \cite{Wang2024Heterogeneous} lacks PE with respect to AP antennas. The model-driven FNN in \cite{Liu2022CellfreeModelDriven} based on regularized zero-forcing only satisfies PE with respect to UEs.

Furthermore, unlike cellular systems where a UE is served by a single AP, a cell-free UE can be served by a variable subset of APs. Consequently, the precoding policy depends on both channel state information and the association status. Existing learning methods for cell-free systems often assume that a UE is associated with all APs and perform joint optimization of precoding and association based on instantaneous channel state information~\cite{Aggarwal2023CellfreeDNN, Chen2024SUNet, Liu2022CellfreeModelDriven, Wang2024Heterogeneous}. These approaches lead to association decisions that fluctuate with small-scale channels, potentially causing significant signaling overhead due to frequent handovers in real-world deployments. In contrast, practical systems typically determine the association based on large-scale channel gains to reduce this overhead. Thus, existing learning methods are unsuitable for scenarios with flexible UE-AP associations.

In this paper, we propose an association-aware GNN (AAGNN) that satisfies the 3D-PE property of the precoding policy in cell-free systems. To capture the relationship between the precoding policy and UE-AP associations, we first formulate the sum-rate maximization problem with inputs that explicitly reflect the association status. We then derive a parameter-sharing structure for AAGNN weight matrix to guarantee the 3D-PE property, based on which the update equation and graph for AAGNN are developed. Finally, we design an attention mechanism to enhance the generalizability across varying numbers of UEs. Simulation results demonstrate the superiority of the AAGNN in learning and generalization performance with low sample, time, and space complexity.
\vspace{-6mm}

\section{System Model}
Consider a cell-free multi-user multi-antenna system where $M$ APs, each with $N$ antennas, serve $K$ randomly located single-antenna UEs. Let $\mathbf{D} \in \left\{ 0,1 \right\} ^ {K \times M}$ denote the UE-AP association matrix, with $d_{km}$ being the element at the $k$-th row and $m$-th column, where $d_{km} = 1$ if AP$_m$ serves UE$_k$ and $d_{km} = 0$ otherwise. Let $\mathbf{h}_{km} \in \mathbb{C}^{N \times 1}$ and $\mathbf{v}_{km} \in \mathbb{C}^{N \times 1}$ denote the channel vector and the precoding vector from AP$_m$ to UE$_k$. Define $\mathbf{H}_m = \left[ \mathbf{h}_{1m},\ldots,\mathbf{h}_{Km} \right] \in \mathbb{C} ^ {N \times K}$ and $\mathbf{V}_m = \left[ \mathbf{v}_{1m},\ldots,\mathbf{v}_{Km} \right] \in \mathbb{C} ^ {N \times K}$ as the channel and precoding matrices of AP$_m$. The channel and precoding matrices from all APs to all UEs are expressed as $\mathbf{H} = \left[ \mathbf{H}_1 ^ \T, \ldots, \mathbf{H}_M ^ \T \right] ^ \T \in \mathbb{C} ^ {NM \times K}$ and $\mathbf{V} = \left[ \mathbf{V}_1 ^ \T, \ldots, \mathbf{V}_M ^ \T \right] ^ \T \in \mathbb{C} ^ {NM \times K}$. 

The precoder optimization problem, aimed at maximizing the sum rate of UEs, subject to per‑AP power constraints, can be formulated as 
\begin{subequations}\label{SMR_problem}
    \vspace{-2mm}
    \begin{align}
        & \max\limits_{\mathbf{V}} \text{  }\sum\limits_{k = 1}^{K} \log \left( 1 + \gamma_k \right) \label{SMR_problem: obj} \vspace{-3mm}  \\ 
        &\ \text{ s.t.}\text{    }\sum\limits_{k = 1} ^ K{\left\| \mathbf{v}_{km} \right\|_2^2}\le P,m=1,\dots,M,
    \end{align}
    \vspace{-3mm}
\end{subequations}
where
\begin{equation}\label{SINR}
    \gamma_k = \frac{\left| \sum\limits_{m = 1}^M  \mathbf{h}_{km} ^ \H d_{km}\mathbf{v}_{km} \right| ^ 2}{\sum\limits_{i \ne k}\left| {\sum\limits_{m = 1}^M{\mathbf{h}_{km} ^ \H d_{im}\mathbf{v}_{im}}}\right|^2 + \delta ^2}
    \vspace{-2mm}
\end{equation}
is the signal to interference plus noise ratio (SINR) of UE$_k$, $\delta ^2$ is the noise power, and $P$ is the total transmit power of each AP. 

\section{Design of AAGNN}
\subsection{UE-AP Association-Aware Precoding Policy}
It can be observed from \eqref{SINR} that the UE-AP association status $d_{km}$ determines the role of the channel $\mathbf{h}_{km}$. Specifically, if $d_{km} = 1$, $\mathbf{h}_{km}$ represents a desired signal channel; otherwise, it represents an interference channel. To explicitly differentiate them, we define $\tilde{\mathbf{h}}_{km} \triangleq d_{km}\mathbf{h}_{km}$ and $\hat{\mathbf{h}}_{km} \triangleq \left( 1 - d_{km} \right)\mathbf{h}_{km}$. Consequently, $\mathbf{h}_{km}$ and $d_{km}$ can be reconstructed as $\mathbf{h}_{km} = \tilde{\mathbf{h}}_{km} + \hat{\mathbf{h}}_{km}$ and $d_{km} = \frac{\left\| \tilde{\mathbf{h}}_{km} \right\|_2}{\left\| \tilde{\mathbf{h}}_{km} + \hat{\mathbf{h}}_{km} \right\|_2}$. Then, $\gamma_k$ in \eqref{SINR} can be rewritten as
\vspace{-2mm}
\begin{equation}\label{SINR_equivalent}
    \vspace{-0.5mm}
    \gamma_k = \frac{\left| \sum\limits_{m = 1}^M{ \tilde{\mathbf{h}}_{km} ^ \H\mathbf{v}_{km}} \right|^2}{\sum\limits_{i \ne k}{\left| \sum\limits_{m = 1}^M{\left( \tilde{\mathbf{h}}_{km} + \hat{\mathbf{h}}_{km} \right) ^ \H \frac{\left\| \tilde{\mathbf{h}}_{im} \right\|_2}{\left\| \tilde{\mathbf{h}}_{im} + \hat{\mathbf{h}}_{im} \right\|_2}\mathbf{v}_{im}} \right|^2 + \delta^2}}.
\end{equation}

By substituting \eqref{SINR_equivalent} into \eqref{SMR_problem}, the optimal precoding policy can be formulated as the mapping $\mathbf{V}^* = \mathcal{F}\left( \mathbf{H}_{D}, \mathbf{H}_{\bar{D}} \right)$, where $\mathbf{H}_{D} \in \mathbb{C} ^ {NM \times K}$ and $\mathbf{H}_{\bar{D}} \in \mathbb{C} ^ {NM \times K}$ are variants of the channel matrix $\mathbf{H}$, constructed by replacing each element $\mathbf{h}_{km}$ with $\tilde{\mathbf{h}}_{km}$ and $\hat{\mathbf{h}}_{km}$, respectively.

The PE property of the precoding policy $\mathcal{F}\left( \cdot \right)$ is analyzed as follows. We examine how the optimal precoder $\mathbf{V}^*$ transforms when the indices of antennas, APs, and UEs in $\mathbf{H}_{D}$ and $\mathbf{H}_{\bar{D}}$ are permuted. Specifically, we permute the antenna indices of AP$_m$ using a permutation matrix $\mathbf{\Pi}_{N,m} \in \mathbb{R}^{N \times N}$, the AP indices using $\mathbf{\Pi}_M \in \mathbb{R}^{M \times M}$, and the UE indices using $\mathbf{\Pi}_K \in \mathbb{R}^{K \times K}$. Under the 3D permutation, the input matrices $\mathbf{H}_{D}$ and $\mathbf{H}_{\bar{D}}$ become $\mathbf{A}\mathbf{H}_D\mathbf{\Pi}_K^\T$ and $\mathbf{A}\mathbf{H}_{\bar{D}}\mathbf{\Pi}_K^\T$, respectively, where $\mathbf{A} \triangleq \left( \mathbf{\Pi}_M \otimes \mathbf{I}_N \right) \mathrm{diag}\left( \mathbf{\Pi}_{N,1}, \ldots, \mathbf{\Pi}_{N,M} \right)$. By substituting these permuted inputs into problem \eqref{SMR_problem}, it can be verified that the objective function and constraints should hold invariant if the precoder is transformed to $\mathbf{A}\mathbf{V}^*\mathbf{\Pi}_K^\T$. Thus, the precoding policy satisfies the following 3D-PE property.
\begin{equation}\label{permutation_property}
    \begin{aligned}
    \mathbf{A}\mathbf{V}^*\mathbf{\Pi}_K^\T = \mathcal{F}\left( \mathbf{A}\mathbf{H}_D\mathbf{\Pi}_K^\T, \mathbf{A}\mathbf{H}_{\bar{D}}\mathbf{\Pi}_K^\T \right).
    \end{aligned}
    \vspace{-1mm}
\end{equation}
Next, we design the AAGNN to satisfy this 3D-PE property.

\subsection{Parameter-Sharing Structure for the  Weight Matrix}
The PE property of a GNN is achieved by enforcing the property at every layer, which is accomplished by appropriately designing the weight matrix \cite{Maron2019InvariantEquivariantNetworks}. Specifically, each layer applies a linear transformation to the input via a learnable weight matrix that must satisfy a specific parameter-sharing structure to ensure the PE property.

Let $\mathbf{X}^{\left( l \right)} = \left[ \left[ \mathbf{x}_{11}^{\left( l \right)},\ldots,\mathbf{x}_{K1}^{\left( l \right)} \right]^\T,\ldots,\left[ \mathbf{x}_{1M}^{\left( l \right)},\ldots,\mathbf{x}_{KM}^{\left( l \right)} \right]^\T \right]^\T \in \mathbb{C}^{NM \times K}$ denote the output of the $l$-th layer, where $l = 1,\dots,L$ and $L$ is the total number of layers. $\mathbf{x}_{km}^{\left( l \right)} \in \mathbb{C}^{N \times 1}$ is the hidden representation, which corresponds to the precoder $\mathbf{v}_{km}$ in the output layer. Given that the policy $\mathcal{F}\left( \cdot \right)$ takes $\mathbf{H}_D$ and $\mathbf{H}_{\bar{D}}$ as input, we select the input to the $l$-th layer similarly as $\mathbf{X}_D^{\left( l - 1 \right)} \in \mathbb{C}^{NM \times K}$ and $\mathbf{X}_{\bar{D}}^{\left( l - 1 \right)} \in \mathbb{C}^{NM \times K}$, which are constructed by replacing each element $\mathbf{x}_{km}^{\left( l - 1 \right)}$ in $\mathbf{X}^{\left( l - 1 \right)}$ with $d_{km}\mathbf{x}_{km}^{\left( l - 1 \right)}$ and $\left( 1- d_{km} \right)\mathbf{x}_{km}^{\left( l - 1 \right)}$, respectively. Then, the relationship between the output and input of the $l$-th layer can be expressed as
\begin{equation}\label{approximate_optimal_function}
    \overrightarrow{\mathbf{X}}^{\left( l \right)} = \sigma^{\left( l \right)} \left( \psi^{\left( l \right)} \left({\overrightarrow{\mathbf{X}}}_D^{\left( l-1 \right)}, {\overrightarrow{\mathbf{X}}}_{\bar{D}}^{\left( l-1 \right)} \right) \right),
    \vspace{-2mm}
\end{equation}
where $\sigma^{\left( l \right)}\left( \cdot \right)$ is a nonlinear activation function, $\psi^{\left( l \right)}\left( \cdot \right)$ is a linear transformation function, and $\overrightarrow{\mathbf{X}}^{\left( l \right)} = \mathrm{vec}\left( \mathbf{X}^{\left( l \right)} \right) \in \mathbb{C}^{NMK \times 1}$, $\overrightarrow{\mathbf{X}}_D^{\left( l - 1 \right)} = \mathrm{vec}\left( \mathbf{X}_D^{\left( l - 1 \right)} \right) \in \mathbb{C}^{NMK \times 1}$ and $\overrightarrow{\mathbf{X}}_{\bar{D}}^{\left( l - 1 \right)} = \mathrm{vec}\left( \mathbf{X}_{\bar{D}}^{\left( l - 1 \right)} \right) \in \mathbb{C}^{NMK \times 1}$ denote the vectorization of $\mathbf{X}^{\left( l - 1 \right)}$, $\mathbf{X}_D^{\left( l - 1 \right)}$ and $\mathbf{X}_{\bar{D}}^{\left( l - 1 \right)}$, respectively.

\begin{figure}
	\centering
	\includegraphics[width=0.650\linewidth]{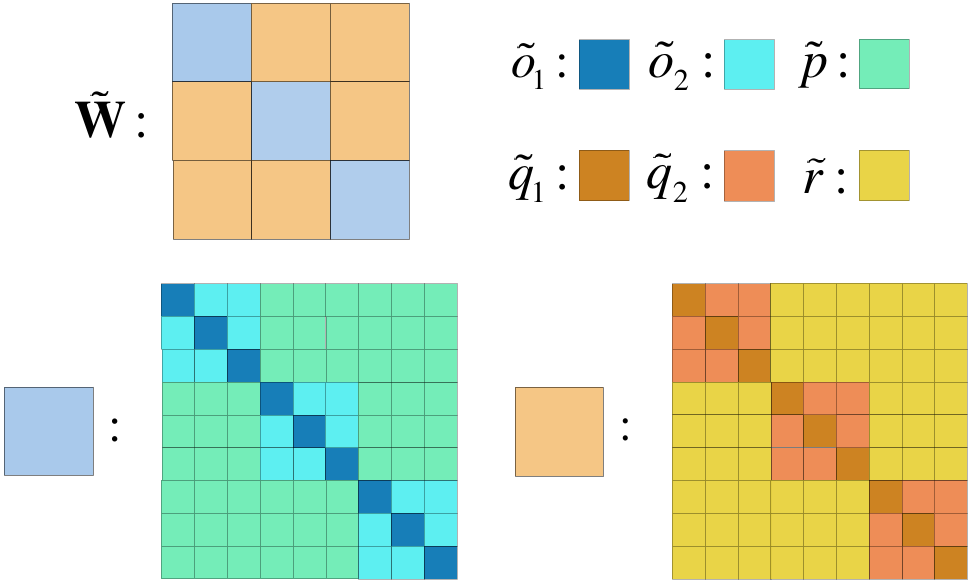}
	\vspace{-1mm}
        \caption{Parameter-sharing structure of $\tilde{\mathbf{W}}$ ($\hat{\mathbf{W}}$ has the same structure).}
	\label{fig: weight_matrix1}
	\vspace{-5mm}
\end{figure}

To ensure the 3D-PE property of each layer, both $\sigma^{\left( l \right)}\left( \cdot \right)$ and $\psi^{\left( l \right)}\left( \cdot \right)$ need to satisfy the property \cite{Maron2019InvariantEquivariantNetworks}. Since $\sigma^{\left( l \right)}\left( \cdot \right)$ is typically a point-wise function that naturally satisfies the 3D-PE property, we focus on designing $\psi^{\left( l \right)}\left( \cdot \right)$, which has the following form
\vspace{-1mm}
\begin{equation}\label{weight_function2}
     \psi^{\left( l \right)} \left({\overrightarrow{\mathbf{X}}}_D^{\left( l-1 \right)}, {\overrightarrow{\mathbf{X}}}_{\bar{D}}^{\left( l-1 \right)} \right) = \mathbf{W}_{\psi}^{\left( l \right)}\left[ \left( \overrightarrow{\mathbf{X}}_D^{\left( l - 1 \right)} \right)^\T \; \left( \overrightarrow{\mathbf{X}}_{\bar{D}}^{\left( l - 1 \right)} \right)^\T \right]^\T,
\end{equation}
where $\mathbf{W}_{\psi}^{\left( l \right)} \in \mathbb{C}^{NMK \times 2NMK}$ is the weight matrix. $\mathbf{W}_{\psi}^{\left( l \right)}$ can be partitioned into two sub-matrices as $\mathbf{W}_{\psi}^{\left( l \right)} = \left[ \tilde{\mathbf{W}} \; \hat{\mathbf{W}} \right]$, where $\tilde{\mathbf{W}},\hat{\mathbf{W}} \in \mathbb{C}^{NMK \times NMK}$ (we omit their subscripts $\left( l \right)$ for simplicity). Then, \eqref{weight_function2} can be rewritten as
\begin{equation}\label{weight_function3}
    \psi^{\left( l \right)} \left({\overrightarrow{\mathbf{X}}}_D^{\left( l-1 \right)}, {\overrightarrow{\mathbf{X}}}_{\bar{D}}^{\left( l-1 \right)} \right) = \tilde{\mathbf{W}}\overrightarrow{\mathbf{X}}_D^{\left( l - 1 \right)} + \hat{\mathbf{W}}\overrightarrow{\mathbf{X}}_{\bar{D}}^{\left( l - 1 \right)}.
    \vspace{-2mm}
\end{equation}
According to the 3D-PE property defined in \eqref{permutation_property}, $\tilde{\mathbf{W}}$ and $\hat{\mathbf{W}}$ must satisfy the following condition
\begin{equation}\label{weight_function3_permutation_property}
    \begin{aligned}
        &\left( \mathbf{\Pi}_K \otimes \mathbf{A} \right)\psi^{\left( l \right)} \left({\overrightarrow{\mathbf{X}}}_D^{\left( l-1 \right)}, {\overrightarrow{\mathbf{X}}}_{\bar{D}}^{\left( l-1 \right)} \right) \\
        &= \tilde{\mathbf{W}}\left( \mathbf{\Pi}_K \otimes \mathbf{A} \right)\overrightarrow{\mathbf{X}}_D^{\left( l - 1 \right)} + \hat{\mathbf{W}}\left( \mathbf{\Pi}_K \otimes \mathbf{A} \right)\overrightarrow{\mathbf{X}}_{\bar{D}}^{\left( l - 1 \right)}.
    \end{aligned}
\end{equation}
Substituting \eqref{weight_function3} into the left-hand side of \eqref{weight_function3_permutation_property} yields
\begin{equation}\label{weight_function3_permutation_property2}
    \begin{aligned}
        &\left( \mathbf{\Pi}_K \otimes \mathbf{A} \right)\left( \tilde{\mathbf{W}}\overrightarrow{\mathbf{X}}_D^{\left( l - 1 \right)} + \hat{\mathbf{W}}\overrightarrow{\mathbf{X}}_{\bar{D}}^{\left( l - 1 \right)} \right) \\
        &= \tilde{\mathbf{W}}\left( \mathbf{\Pi}_K \otimes \mathbf{A} \right)\overrightarrow{\mathbf{X}}_D^{\left( l - 1 \right)} + \hat{\mathbf{W}}\left( \mathbf{\Pi}_K \otimes \mathbf{A} \right)\overrightarrow{\mathbf{X}}_{\bar{D}}^{\left( l - 1 \right)}.
        \vspace{-2mm}
    \end{aligned}
\end{equation}
The condition in \eqref{weight_function3_permutation_property2} must hold for arbitrary $\overrightarrow{\mathbf{X}}_D^{\left( l - 1 \right)}$ and $\overrightarrow{\mathbf{X}}_{\bar{D}}^{\left( l - 1 \right)}$. To characterize the structures of $\tilde{\mathbf{W}}$, we consider a special case where $\overrightarrow{\mathbf{X}}_{\bar{D}}^{\left( l - 1 \right)} = \mathbf{0}$. In this case, \eqref{weight_function3_permutation_property2} reduces to
\begin{equation}
    \left( \mathbf{\Pi}_K \otimes \mathbf{A} \right)\tilde{\mathbf{W}}\overrightarrow{\mathbf{X}}_D^{\left( l - 1 \right)} = \tilde{\mathbf{W}}\left( \mathbf{\Pi}_K \otimes \mathbf{A} \right)\overrightarrow{\mathbf{X}}_D^{\left( l - 1 \right)}.
\end{equation}
To make this condition hold for arbitrary $\overrightarrow{\mathbf{X}}_D^{\left( l - 1 \right)}$, $\tilde{\mathbf{W}}$ needs to satisfy the condition
\begin{equation}\label{W1_permutation_property2}    
    \left( \mathbf{\Pi}_K \otimes \mathbf{A} \right) \tilde{\mathbf{W}} = \tilde{\mathbf{W}}\left( \mathbf{\Pi}_K \otimes \mathbf{A} \right).
\end{equation}
Similarly, by setting $\overrightarrow{\mathbf{X}}_D^{\left( l - 1 \right)} = \mathbf{0}$, we can obtain an analogous condition for $\hat{\mathbf{W}}$, which is
\begin{equation}\label{W2_permutation_property2}    
    \left( \mathbf{\Pi}_K \otimes \mathbf{A} \right) \hat{\mathbf{W}} = \hat{\mathbf{W}}\left( \mathbf{\Pi}_K \otimes \mathbf{A} \right).
\end{equation}
It is evident that if \eqref{W1_permutation_property2} and \eqref{W2_permutation_property2} hold, the condition in \eqref{weight_function3_permutation_property2} holds. Thus, these two conditions are necessary and sufficient for ensuring the 3D-PE property. The conditions \eqref{W1_permutation_property2} and \eqref{W2_permutation_property2} must hold for arbitrary permutation matrices $\mathbf{\Pi}_M$, $\mathbf{\Pi}_K$ and $\mathbf{\Pi}_{N,m}, m = 1,\dots,M$. This imposes a strict parameter-sharing structure on $\tilde{\mathbf{W}}$ and $\hat{\mathbf{W}}$. Following the methods in \cite{Maron2019InvariantEquivariantNetworks} and \cite{Wang2020HierarchicalPE}, we obtain that $\tilde{\mathbf{W}}$ consists of 6 free parameters, denoted by $\tilde{o}_1$, $\tilde{o}_2$, $\tilde{p}$, $\tilde{q}_1$, $\tilde{q}_2$, and $\tilde{r}$. Similarly, $\hat{\mathbf{W}}$ also consists of 6 free parameters, denoted by $\hat{o}_1$, $\hat{o}_2$, $\hat{p}$, $\hat{q}_1$, $\hat{q}_2$, and $\hat{r}$. The resulting parameter-sharing structure is illustrated in Fig. \ref{fig: weight_matrix1}.

By substituting the structures of $\tilde{\mathbf{W}}$ and $\hat{\mathbf{W}}$ into \eqref{weight_function2} and then into \eqref{approximate_optimal_function}, we can derive that $x_{kmn}^{\left( l \right)}$, the $n$-th element of $\mathbf{x}^{\left( l \right)}_{km}$, is updated by
\begin{equation}\label{update_equation1}
    \begin{aligned}
        & x_{kmn}^{\left( l \right)} = \sigma^{\left( l \right)}\Big(\tilde{o}_1d_{km}x_{kmn}^{\left( l - 1 \right)} + \hat{o}_1\left( 1 - d_{km} \right)x_{kmn}^{\left( l - 1 \right)} + \\
        & \tilde{o}_2\sum\limits_{c \ne n}{d_{km}x_{kmc}^{\left( l - 1 \right)}} + \hat{o}_2\sum\limits_{c \ne n}{\left( 1 - d_{km} \right)x_{kmc}^{\left( l - 1 \right)}} + \\
        & \tilde{p}\sum\limits_{b \ne m}{\sum\limits_{c}{d_{kb}x_{kbc}^{\left( l - 1 \right)}}} + \hat{p}\sum\limits_{b \ne m}{\sum\limits_{c}{\left( 1 - d_{kb} \right)x_{kbc}^{\left( l - 1 \right)}}} + \\
        & \tilde{q}_1\sum\limits_{a \ne k}{d_{am}x_{amn}^{\left( l - 1 \right)}} + \hat{q}_1\sum\limits_{a \ne k}{\left( 1 - d_{am} \right)x_{amn}^{\left( l - 1 \right)}} + \\
        & \tilde{q}_2\sum\limits_{a \ne k}{\sum\limits_{c \ne n}{d_{am}x_{amc}^{\left( l - 1 \right)}}} + \hat{q}_2\sum\limits_{a \ne k}{\sum\limits_{c \ne n}{\left( 1 - d_{am} \right)x_{amc}^{\left( l - 1 \right)}}} + \\
        & \tilde{r}\sum\limits_{a \ne k}{\sum\limits_{b \ne m}{\sum\limits_{c}{d_{ab}x_{abc}^{\left( l - 1 \right)}}}} + \hat{r}\sum\limits_{a \ne k}{\sum\limits_{b \ne m}{\sum\limits_{c}{\left( 1 - d_{ab} \right)x_{abc}^{\left( l - 1 \right)}}}}\Big).
    \end{aligned}
    \vspace{-6mm}
\end{equation}

\subsection{Graph Construction and Update Equation}
\begin{figure}
	\centering
	\includegraphics[width=0.60\linewidth]{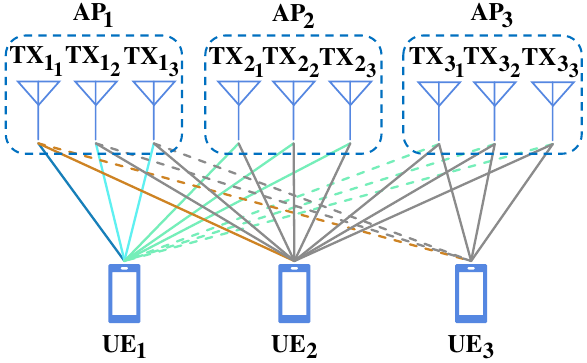}
	\vspace{-1mm}
    \caption{Illustration of the graph of AAGNN with $K = 3$, $N = 3$ and $M = 3$ (TX$_{m_n}$ denotes the $n$-th antenna vertex of AP$_m$).}
	\label{fig: graph}
	\vspace{-4mm}
\end{figure}

To implement the update equation \eqref{update_equation1}, we construct a heterogeneous graph consisting of AP antenna vertices and UE vertices, with edges connecting the two types of vertices. The graph is illustrated in Fig. \ref{fig: graph}, where the edges between antenna vertices of an AP and its served UE vertices are depicted by solid lines, while other edges are shown as dashed lines.

The vertices have no features or actions. For the edge connecting the $k$-th UE vertex and the $n$-th antenna vertex of AP$_m$, its feature is the $n$-th element of $\mathbf{h}_{km}$, denoted by $h_{kmn}$. At the $l$-th layer, its hidden representation is $x_{kmn}^{\left( l \right)}$, and at the output layer, it generates action $v_{kmn}$, which is the $n$-th element of $\mathbf{v}_{km}$.

For GNNs where features and actions are defined on edges, the update of the hidden representations consists of aggregation and combination steps \cite{Zhao2022EGNN, Guo2021PENN}. Specifically, for the edges connected to the same vertex, their hidden representations of the previous layer (say the $( l - 1 )$-th layer) are first aggregated. Then, each edge combines the aggregated information from its adjacent edges (i.e., edges that share a common vertex) to update its own hidden representation. 
When mapping the update equation \eqref{update_equation1} to these two steps, we find that the terms associated with parameters $\tilde{q}_2$, $\tilde{r}$, $\hat{q}_2$, and $\hat{r}$ correspond to the aggregation from non-adjacent edges. Therefore, to avoid such aggregation, we set parameters $\tilde{q}_2$, $\tilde{r}$, $\hat{q}_2$, and $\hat{r}$ to zero and finally obtain the aggregation and combination steps as follows.
\begin{itemize}
	\item[(i)] \textbf{Aggregation}:
    \begin{equation}\label{update_equation}
        \begin{aligned}
            u_{kmn}^{\left( l \right)} = &\tilde{o}_2\sum\limits_{c \ne n}{d_{km}x_{kmc}^{\left( l - 1 \right)}} + \hat{o}_2\sum\limits_{c \ne n}{\left( 1 - d_{km} \right)x_{kmc}^{\left( l - 1 \right)}} + \\
            & \tilde{p}\sum\limits_{b \ne m}{\sum\limits_{c}{d_{kb}x_{kbc}^{\left( l - 1 \right)}}} + \hat{p}\sum\limits_{b \ne m}{\sum\limits_{c}{\left( 1 - d_{kb} \right)x_{kbc}^{\left( l - 1 \right)}}}, \\
            w_{kmn}^{\left( l \right)} = & \tilde{q}_1\sum\limits_{a \ne k}{d_{am}x_{amn}^{\left( l - 1 \right)}} + \hat{q}_1\sum\limits_{a \ne k}{\left( 1 - d_{am} \right)x_{amn}^{\left( l - 1 \right)}}.
        \end{aligned}
        \vspace{-2mm}
    \end{equation}
    \item[(ii)] \textbf{Combination}:
    \begin{equation}\label{update_equation}
        \begin{aligned}
            x_{kmn}^{\left( l \right)} = &\sigma^{\left( l \right)}\Big(\tilde{o}_1d_{km}x_{kmn}^{\left( l - 1 \right)} + \hat{o}_1\left( 1 - d_{km} \right)x_{kmn}^{\left( l - 1 \right)} + \\
            & u_{kmn}^{\left( l \right)} + w_{kmn}^{\left( l \right)} \Big).
        \end{aligned}
        \vspace{-6mm}
    \end{equation}
\end{itemize}
\subsection{Design of Attention Mechanism}
\begin{figure}[!htb]
    \vspace{-3mm}
	\centering
	\includegraphics[width=0.650\linewidth]{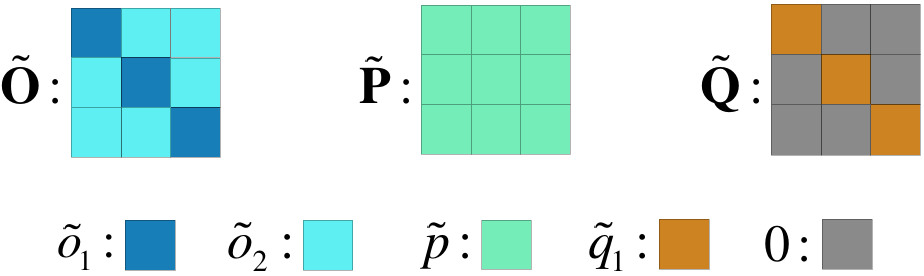}
        \caption{Parameter-sharing structures of $\tilde{\mathbf{O}}$, $\tilde{\mathbf{P}}$ and $\tilde{\mathbf{Q}}$ ($\hat{\mathbf{O}}$, $\hat{\mathbf{P}}$ and $\hat{\mathbf{Q}}$ have the same structures).}
	\label{fig: weight_matrix4}
	\vspace{-3mm}
\end{figure}

Attention mechanism can enhance the generalization performance of precoding to the number of UEs in cellular systems~\cite{Li2024GNN}. In this subsection, we develop a specialized attention mechanism for AAGNN in cell-free systems.

First, by expressing the representations of all edges connected to the antenna vertices of the same AP$_m$ as a vector $\mathbf{x}_{km}^{\left( l \right)}$, \eqref{update_equation} can be rewritten as
\begin{equation}\label{update_equation_vector}
    \begin{aligned}
        \mathbf{x}_{km}^{\left( l \right)} = & \tilde{\mathbf{O}}d_{km}\mathbf{x}_{km}^{\left( l - 1 \right)} + \hat{\mathbf{O}}\left( 1 - d_{km} \right)\mathbf{x}_{km}^{\left( l - 1 \right)} + \\
        & \tilde{\mathbf{P}}\sum\limits_{b \ne m}{d_{kb}\mathbf{x}_{kb}^{\left( l - 1 \right)}} + \hat{\mathbf{P}}\sum\limits_{b \ne m}{\left( 1 - d_{kb} \right)\mathbf{x}_{kb}^{\left( l - 1 \right)}} + \\
        & \tilde{\mathbf{Q}}\sum\limits_{a \ne k}{d_{am}\mathbf{x}_{am}^{\left( l - 1 \right)}} + \hat{\mathbf{Q}}\sum\limits_{a \ne k}{\left( 1 - d_{am} \right)\mathbf{x}_{am}^{\left( l - 1 \right)}},
    \end{aligned}
    \vspace{-2mm}
\end{equation}
where the parameter-sharing structures of the matrices $\tilde{\mathbf{O}}$, $\tilde{\mathbf{P}}$, $\tilde{\mathbf{Q}}$, $\hat{\mathbf{O}}$, $\hat{\mathbf{P}}$, and $\hat{\mathbf{Q}}$ are illustrated in Fig.~\ref{fig: weight_matrix4}. To facilitate the attention design, we decompose \eqref{update_equation_vector} as
\begin{equation}\label{update_equation_vector2}
    \mathbf{x}_{km}^{\left( l \right)} = \mathbf{t}_{km}^{\left( l - 1 \right)} + \sum\limits_{a \ne k}{\mathbf{z}_{am}^{\left( l - 1 \right)}},
    \vspace{-3mm}
\end{equation}
where $\mathbf{t}_{km}^{\left( l - 1 \right)}$ and $\mathbf{z}_{am}^{\left( l - 1 \right)}$ are defined as 
\vspace{-1.5mm}
\begin{subequations}\label{t_z_defination}
    \begin{align}
        \begin{split}
            \mathbf{t}_{km}^{\left( l - 1 \right)} = & \tilde{\mathbf{O}}d_{km}\mathbf{x}_{km}^{\left( l - 1 \right)} + \hat{\mathbf{O}}\left( 1 - d_{km} \right)\mathbf{x}_{km}^{\left( l - 1 \right)} + \\
            & \tilde{\mathbf{P}}\sum\limits_{b \ne m}{d_{kb}\mathbf{x}_{kb}^{\left( l - 1 \right)}} + \hat{\mathbf{P}}\sum\limits_{b \ne m}{\left( 1 - d_{kb} \right)\mathbf{x}_{kb}^{\left( l - 1 \right)}},
        \end{split} \\
        \mathbf{z}_{am}^{\left( l - 1 \right)} = & \tilde{\mathbf{Q}}d_{am}\mathbf{x}_{am}^{\left( l - 1 \right)} + \hat{\mathbf{Q}}\left( 1 - d_{am} \right)\mathbf{x}_{am}^{\left( l - 1 \right)}.
        \vspace{-0.5mm}
    \end{align}
\end{subequations}
It can be observed that $\mathbf{t}_{km}^{\left( l - 1 \right)}$ represents the features from the edges themselves and the edges connected to the same UE vertices, while $\mathbf{z}_{am}^{\left( l - 1 \right)}$ represents the features from the edges connected to the same antenna vertices. Instead of directly combining them as in \eqref{update_equation_vector2}, we introduce attention coefficients to weight $\mathbf{t}_{km}^{\left( l - 1 \right)}$ and $\mathbf{z}_{am}^{\left( l - 1 \right)}$, leading to the following update equation
\begin{equation}\label{update_equation_vector_attention1}
    \vspace{-2.5mm}
    \mathbf{x}_{km}^{\left( l \right)} = \lambda_{kkm}^{\left( l - 1 \right)}\mathbf{t}_{km}^{\left( l - 1 \right)} + \sum\limits_{a \ne k}{\lambda_{akm}^{\left( l - 1 \right)}\mathbf{z}_{am}^{\left( l - 1 \right)}},
    \vspace{-1.5mm}
\end{equation}
where $\lambda_{kkm}^{\left( l - 1 \right)}$ and $\lambda_{akm}^{\left( l - 1 \right)}$ are scalar attention coefficients reflecting the importance of $\mathbf{t}_{km}^{\left( l - 1 \right)}$ and $\mathbf{z}_{am}^{\left( l - 1 \right)}$ for the update of $\mathbf{x}_{km}^{\left( l \right)}$.

Since the output $\mathbf{x}_{km}^{\left( L \right)}$ (the precoder) should have a high correlation with the channel $\mathbf{h}_{km}$ to enhance signal power, we define the attention coefficients based on the correlation between $\mathbf{h}_{km}$ and $\mathbf{t}_{km}^{\left( l - 1 \right)}$ or $\mathbf{z}_{am}^{\left( l - 1 \right)}$. Specifically, we set $\lambda_{kkm}^{\left( l - 1 \right)} = \alpha^{\left( l \right)}\mathbf{h}_{km}^\H \mathbf{t}_{km}^{\left( l - 1 \right)}$ and $\lambda_{akm}^{\left( l - 1 \right)} = \beta^{\left( l \right)}\mathbf{h}_{km}^\H \mathbf{z}_{am}^{\left( l - 1 \right)}$, where $\alpha^{\left( l \right)}$ and $\beta^{\left( l \right)}$ are learnable parameters. Substituting these into~\eqref{update_equation_vector_attention1} yields the final update equation for AAGNN as
\begin{equation}\label{update_equation_vector_attention2}
    \mathbf{x}_{km}^{\left( l \right)} = \big( \alpha^{\left( l \right)}\mathbf{h}_{km}^\H \mathbf{t}_{km}^{\left( l - 1 \right)} \big)\mathbf{t}_{km}^{\left( l - 1 \right)} + \sum\limits_{a \ne k}{\big( \beta^{\left( l \right)}\mathbf{h}_{km}^\H \mathbf{z}_{am}^{\left( l - 1 \right)} \big)\mathbf{z}_{am}^{\left( l - 1 \right)}}.
\end{equation}
\vspace{-0.5mm}
It is straightforward to verify that the update equation in \eqref{update_equation_vector_attention2} satisfies the 3D-PE property in \eqref{permutation_property}.

\section{SIMULATION RESULTS}
In this section, we evaluate the learning performance, generalization performance, and complexity of AAGNN by comparing it with baseline numerical algorithms and learning-based methods. The compared methods are as follows.
\begin{itemize}
    \item \textbf{WMMSE}: This is a near-optimal numerical algorithm for solving the sum-rate maximization problem \cite{Shi2011WMMSE}.
    \item \textbf{AAGNN-woA}: This is a variant of the proposed GNN without the attention mechanism.
    \item \textbf{EGAT}: This GNN applies the attention mechanism in \cite{Wang2021EGAT} to the above AAGNN-woA.
    \item \textbf{SUNet}: This is the CNN proposed in \cite{Chen2024SUNet} to solve the sum-rate maximization problem in cell-free systems.
    \item \textbf{GNN in \cite{Wang2024Learning}}: This is the GNN proposed in \cite{Wang2024Learning} to solve the sum-rate maximization problem in cell-free systems.
    \item \textbf{MRT$+$GNN}: This is the method proposed in \cite{Raghunath2024MRTGNN} for precoding design in cell-free systems, which utilizes maximum ratio transmission (MRT) beamforming vectors and employs a GNN to learn the power allocation.
\end{itemize}

In simulations, all APs are placed within a square with an inter-site distance of 400~m. UEs are randomly located from the union of discs with a radius of $400/\sqrt{3}$~m centered at the APs. Each UE is served by the set of APs located within a radius of 300~m. The small-scale fading follows the Rayleigh distribution, and the large-scale fading (dB) follows the 3GPP specification \cite{3gppTR38901}, given by $\mathrm{PL} = 13.54 + 39.08\log_{10}\left( d_{\mathrm{3D}} \right)$, where $d_{\mathrm{3D}}$ is the UE-AP distance in meters. Unless otherwise specified, the signal-to-noise ratio (SNR) at the cell edge is set to 10~dB.

All learning-based methods are trained via unsupervised learning, with a batch size of 64 and a learning rate of 0.005. The training dataset consists of 12,800 samples, and the test dataset consists of 1,000 samples. To evaluate learning and generalization performance, the sum rate achieved by all methods is normalized by the sum rate achieved by WMMSE. All results are obtained on a personal computer with an Intel Core i9-13900KF CPU and an NVIDIA RTX 4080 GPU.

\subsection{Performance Comparison}
\begin{figure}[htb!]
    \vspace{-4mm}
	\centering
	\subfigure[$K = 8$]{%
    \begin{minipage}[c]{0.509\linewidth}
        \centering
        \includegraphics[width=1\linewidth]{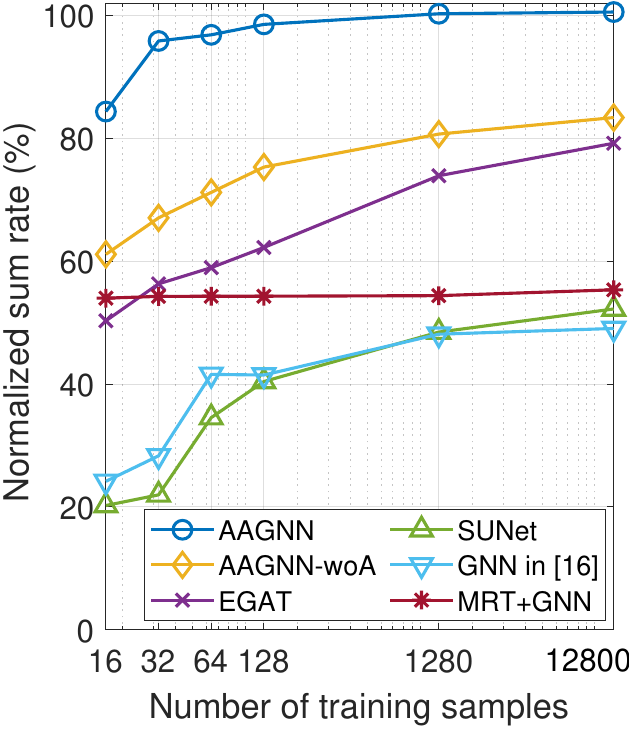}
    \end{minipage}
	}%
	\subfigure[$K = 16$]{%
    \begin{minipage}[c]{0.481\linewidth}
        \centering
        \includegraphics[width=1\linewidth]{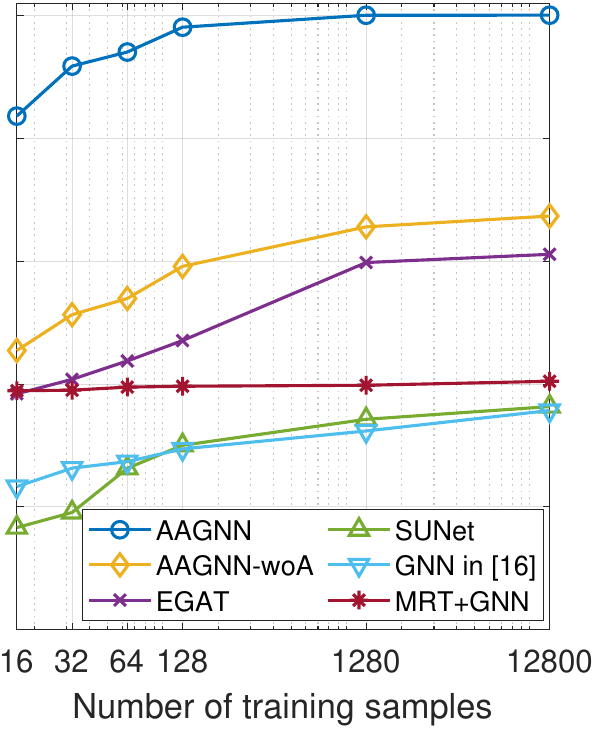}
    \end{minipage}
	}%
	\vspace{-3mm}
    \caption{Learning performance vs. the number of training samples with $N = 16$ and $M = 4$.}
	\label{fig: learning_performance}
	\vspace{-3mm}
\end{figure}

In Fig. \ref{fig: learning_performance}, the learning performance is compared under different numbers of training samples, where $K = 8$ and $K = 16$ are considered in two subfigures, respectively. It can be observed that as the number of training samples decreases, all methods exhibit performance degradation, while AAGNN consistently achieves the best learning performance. Even with only 16 training samples, AAGNN can maintain performance over $80\%$.

\begin{figure}[htb!]
	\centering
	\subfigure[]{%
    \label{fig: generalization_performance_ue}
    \begin{minipage}[c]{0.541\linewidth}
        \centering
        \includegraphics[width=1\linewidth]{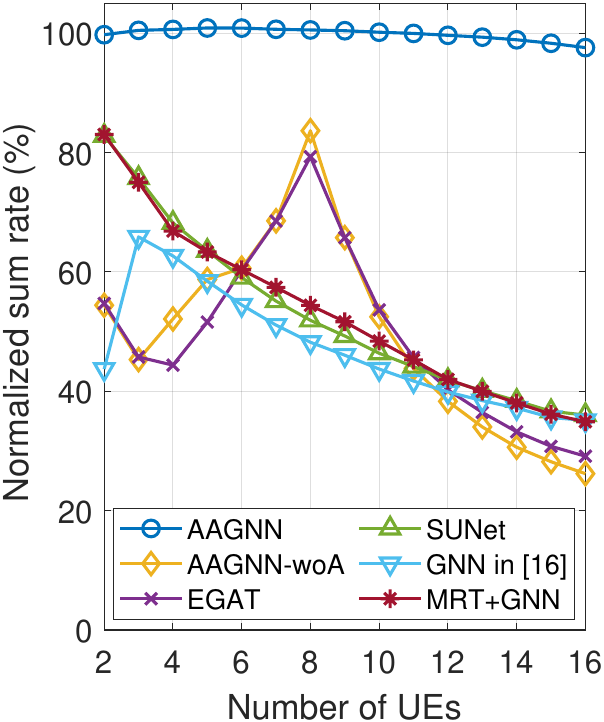}
    \end{minipage}
	}%
	\subfigure[]{%
    \label{fig: generalization_performance_tx}
    \begin{minipage}[c]{0.449\linewidth}
        \centering
        \includegraphics[width=1\linewidth]{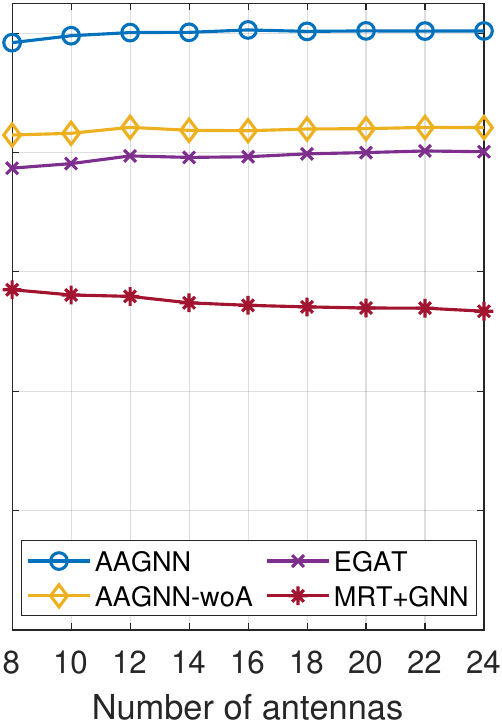}
    \end{minipage}
	}%
	\vspace{-2mm}
    \caption{Generalization performance: (a) generalization to the number of UEs with $N = 16$ and $M = 4$, and (b) generalization to the number of antennas with $K = 8$ and $M = 4$.}
	\label{fig: generalization_performance_ue_tx}
	\vspace{-2mm}
\end{figure}

\begin{table*}
    \centering
    \caption{Training Complexity and Inference Time.}\label{table: complexity}
    \footnotesize
    \begin{threeparttable}
    \begin{tabular}{c|c|c|c|c|c|c}
            \hline\hline
               & AAGNN & AAGNN-woA & EGAT & SUNet & GNN in \cite{Wang2024Learning} & MRT$+$GNN \\
            \hline
            Sample complexity & 12 & 500 & 900 & $>$12800${}^{*}$ & $>$12800${}^{*}$ & $>$12800${}^{*}$ \\
            \hline
            Time complexity (GPU) & 6.683s & 44.735s & 181.74s & $>$303.993s & $>$212.304s & $>$419.969s \\
            \hline
            Space complexity  & 115617 & 397313 & 399365 & $>$444160 & $>$776545 & $>$305528 \\
            \hline
            Inference time (GPU) & 2.432ms & 1.806ms & 1.973ms & $>$2.236ms & $>$1.139ms & $>$5.242ms \\
            \hline
            Inference time (CPU) & 3.731ms & 6.767ms & 9.077ms & $>$4.082ms & $>$1.815ms & $>$8.681ms \\
            \hline\hline
        \end{tabular}
        \begin{tablenotes}
        \item ${}^{*}$: For the case marked by ${}^{*}$, the DNN cannot achieve the target performance of $90\%$ with $12800$ training samples. For these DNNs, we record the time complexity to complete 100 training epochs.
        \end{tablenotes}
    \vspace{-5mm}
    \end{threeparttable}
\end{table*}

In Fig. \ref{fig: generalization_performance_ue}, we compare the generalization performance to the number of UEs. All methods are trained with $K = 8$ and tested for $K \in \left[ 2,16 \right]$. It can be seen that AAGNN generalizes well, achieving performance close to WMMSE across the varying number of UEs, whereas AAGNN-woA and EGAT exhibit significant performance degradation.

\begin{figure}
	\centering
	\includegraphics[width=0.90\linewidth]{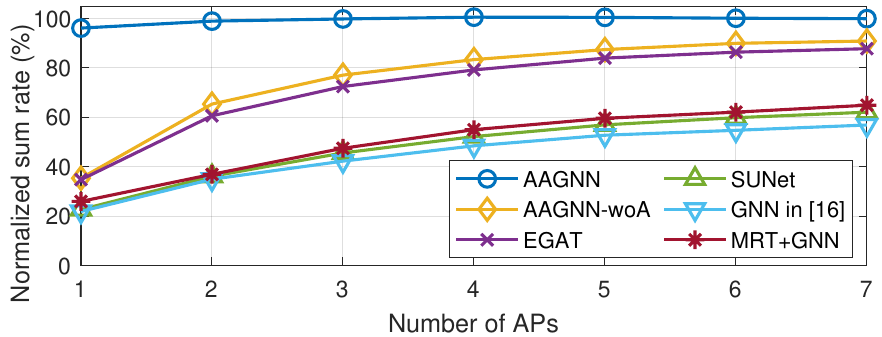}
	\vspace{-1mm}
        \caption{Generalization performance to the number of APs with $K = 8$ and $N = 16$.}
	\label{fig: generalization_performance_ap}
	\vspace{-6mm}
\end{figure}

In Fig. \ref{fig: generalization_performance_tx}, we compare the generalization performance to the number of AP antennas, where SUNet and GNN in \cite{Wang2024Learning} are not included because they cannot generalize to the number of AP antennas. All methods are trained with $N = 16$ and tested for $N \in \left[ 8,24 \right]$. It is shown that all methods exhibit good generalizability to the number of AP antennas.

In Fig. \ref{fig: generalization_performance_ap}, we compare the generalization performance to the number of APs. All methods are trained with $M = 4$ and tested for $M \in \left[ 1,7 \right]$. AAGNN maintains performance close to WMMSE across the varying number of APs, while other methods suffer from performance degradation as the number of APs decreases.

\subsection{Complexity Comparison}
In Table \ref{table: complexity}, we evaluate the sample, time, and space complexity of all DNNs, which are defined respectively as the minimal number of training samples, the time used for training, and the number of free parameters required to achieve the target performance. The target performance is set to 90$\%$ of WMMSE. To ensure that multiple learning methods attain this target, we consider the setup with $K = 4$, $N = 16$, $M = 4$, and $\mathrm{SNR} = 5\mathrm{~dB}$. Among the evaluated methods, only AAGNN, AAGNN-woA, and EGAT can meet the target performance. AAGNN has much lower sample, time, and space complexity compared to AAGNN-woA and EGAT while maintaining a comparable inference time.

\section{CONCLUSION}
In this paper, we proposed the AAGNN for precoder learning in cell-free systems, which can explicitly adapt to the dynamic association between UEs and APs. The proposed architecture satisfies the 3D-PE property that significantly reduces the training complexity, and incorporates a tailored attention mechanism that effectively enhances the generalization performance to the number of UEs. Simulation results demonstrate that AAGNN achieves better learning and generalization performance than baseline learning methods, with much lower sample, time, and space complexity.

\bibliography{AAGNN_20251203}

\begin{thebibliography}{10}
\providecommand{\url}[1]{#1}
\csname url@samestyle\endcsname
\providecommand{\newblock}{\relax}
\providecommand{\bibinfo}[2]{#2}
\providecommand{\BIBentrySTDinterwordspacing}{\spaceskip=0pt\relax}
\providecommand{\BIBentryALTinterwordstretchfactor}{4}
\providecommand{\BIBentryALTinterwordspacing}{\spaceskip=\fontdimen2\font plus
\BIBentryALTinterwordstretchfactor\fontdimen3\font minus \fontdimen4\font\relax}
\providecommand{\BIBforeignlanguage}[2]{{%
\expandafter\ifx\csname l@#1\endcsname\relax
\typeout{** WARNING: IEEEtran.bst: No hyphenation pattern has been}%
\typeout{** loaded for the language `#1'. Using the pattern for}%
\typeout{** the default language instead.}%
\else
\language=\csname l@#1\endcsname
\fi
#2}}
\providecommand{\BIBdecl}{\relax}
\BIBdecl

\bibitem{Demir2021Cellfree}
{\"O}.~T. Demir, E.~Bj{\"o}rnson, and L.~Sanguinetti, ``{Foundations of User-Centric Cell-Free Massive MIMO},'' \emph{Found. Trends Signal Process.}, vol.~14, no. 3-4, pp. 162--472, 2021.

\bibitem{Shi2018DNN}
H.~Sun, X.~Chen, Q.~Shi, M.~Hong, X.~Fu, and N.~D. Sidiropoulos, ``{Learning to Optimize: Training Deep Neural Networks for Interference Management},'' \emph{IEEE Trans. Signal Process.}, vol.~66, no.~20, pp. 5438--5453, Oct. 2018.

\bibitem{Kim2020LearningMethod}
J.~Kim, H.~Lee, S.-E. Hong, and S.-H. Park, ``{Deep Learning Methods for Universal MISO Beamforming},'' \emph{IEEE Wireless Commun. Lett.}, vol.~9, no.~11, pp. 1894--1898, Nov. 2020.

\bibitem{Bronstein2021InductiveBiase}
M.~M. Bronstein, J.~Bruna, T.~Cohen, and P.~Veli{\v{c}}kovi{\'c}, ``{Geometric Deep Learning: Grids, Groups, Graphs, Geodesics, and Gauges},'' \emph{arXiv:2104.13478}, 2021.

\bibitem{Zhao2022EGNN}
B.~Zhao, J.~Guo, and C.~Yang, ``{Learning Precoding Policy: CNN or GNN?}'' \emph{IEEE WCNC}, 2022.

\bibitem{Aggarwal2023CellfreeDNN}
M.~Aggarwal, S.~Deshpande, P.~Sharma, and S.~Ahuja, ``{Beyond 5G: Exploiting Learning Aided Precoder for Downlink Cell-Free Networks},'' \emph{Authorea Preprints}, 2023.

\bibitem{Chen2024SUNet}
G.~Chen, Z.~Wang, Y.~Jia, Y.~Huang, and L.~Yang, ``{An Efficient Architecture Search for Scalable Beamforming Design in Cell-Free Systems},'' \emph{IEEE Trans. Veh. Technol.}, vol.~73, no.~7, pp. 10\,241--10\,253, July 2024.

\bibitem{Wang2024Heterogeneous}
Z.~Wang and V.~Wong, ``{Heterogeneous Graph Neural Network for Cooperative ISAC Beamforming in Cell-Free MIMO Systems},'' \emph{MobiCom}, 2024.

\bibitem{Liu2022CellfreeModelDriven}
S.~Liu, Z.~Gao, C.~Hu, S.~Tan, L.~Fang, and L.~Qiao, ``{Model-Driven Deep Learning Based Precoding for FDD Cell-Free Massive MIMO with Imperfect CSI},'' \emph{IEEE IWCMC}, 2022.

\bibitem{Maron2019InvariantEquivariantNetworks}
H.~Maron, H.~Ben-Hamu, N.~Shamir, and Y.~Lipman, ``{Invariant and Equivariant Graph Networks},'' \emph{ICLR}, 2019.

\bibitem{Wang2020HierarchicalPE}
R.~Wang, M.~Albooyeh, and S.~Ravanbakhsh, ``{Equivariant Networks for Hierarchical Structures},'' \emph{NeurIPS}, 2020.

\bibitem{Guo2021PENN}
J.~Guo and C.~Yang, ``{Learning Power Allocation for Multi-Cell-Multi-User Systems with Heterogeneous Graph Neural Networks},'' \emph{IEEE Trans. Wireless Commun.}, vol.~21, no.~2, pp. 884--897, Feb. 2021.

\bibitem{Li2024GNN}
Y.~Li, Y.~Lu, B.~Ai, O.~A. Dobre, Z.~Ding, and D.~Niyato, ``{GNN-Based Beamforming for Sum-Rate Maximization in MU-MISO Networks},'' \emph{IEEE Trans. Wireless Commun.}, vol.~23, no.~8, pp. 9251--9264, Aug. 2024.

\bibitem{Shi2011WMMSE}
Q.~Shi, M.~Razaviyayn, Z.-Q. Luo, and C.~He, ``{An Iteratively Weighted MMSE Approach to Distributed Sum-Utility Maximization for A MIMO Interfering Broadcast Channel},'' \emph{IEEE Trans. Signal Process.}, vol.~59, no.~9, pp. 4331--4340, Sept. 2011.

\bibitem{Wang2021EGAT}
Z.~Wang, J.~Chen, and H.~Chen, ``{EGAT: Edge-Featured Graph Attention Network},'' \emph{ICANN}, 2021.

\bibitem{Wang2024Learning}
L.~Wang, C.~Chen, J.~Zhang, and C.~Fischione, ``{Learning-Based Joint Antenna Selection and Precoding Design for Cell-Free MIMO Networks},'' \emph{arXiv:2404.08607}, 2024.

\bibitem{Raghunath2024MRTGNN}
R.~Raghunath, B.~Peng, and E.~A. Jorswieck, ``{Energy-Efficient Power Allocation in Cell-Free Massive MIMO via Graph Neural Networks},'' \emph{IEEE ICMLCN}, 2024.

\bibitem{3gppTR38901}
ETSI, ``{Study on Channel Model for Frequencies From 0.5 to 100 GHz},'' 3rd Generation Partnership Project (3GPP), Tech. Rep., 2020.

\end{thebibliography}
\vspace{-1mm}
\end{document}